# Proper Projective Symmetry in some well known Conformally flat Space-Times

Ghulam Shabbir

Faculty of Engineering Sciences

GIK Institute of Engineering Sciences and Technology

Topi Swabi, NWFP, Pakistan

Email: shabbir@giki.edu.pk

**Abstract**

A study of conformally flat but non flat Bianchi type I and cylindrically symmetric static space-times according to proper projective symmetry is given by using some algebraic and direct integration techniques. It is shown that the special class of the above space-times admit proper projective vector fields.

## 1. INTRODUCTION

Through out $M$ is representing the four dimensional, connected, hausdorff space-time manifold with Lorentz metric $g$ of signature (-, +, +, +). The curvature tensor associated with $g_{ab}$, through Levi-Civita connection, is denoted in component form by $R^a{}_{bcd}$, the Weyl tensor components are $C^a{}_{bcd}$, and the Ricci tensor components are $R_{ab} = R^c{}_{acb}$. The usual covariant, partial and Lie derivatives are denoted by a semicolon, a comma and the symbol $L$, respectively. Round and square brackets denote the usual symmetrization and skew-symmetrization, respectively. A space-time is said to be conformally flat if $C^a{}_{bcd} = 0$ every where on $M$. Finally, $M$ is assumed to be non-flat in the sense that the curvature tensor does not vanish over a non-empty open subset of $M$ and is not of constant curvature.

Any vector field $X$ on $M$ can be decomposed as

$$X_{a;b} = \frac{1}{2}h_{ab} + F_{ab}, \qquad (1)$$

where $h_{ab}(=h_{ba}) = L_X g_{ab}$ and $F_{ab}(=-F_{ba})$ are symmetric and skew symmetric tensor on $M$, respectively. Such a vector field $X$ is called projective if the local diffeomorphisms $\psi_t$ (for appropriate $t$) associated with $X$ map geodesics into geodesics. This is equivalent to the condition that $h_{ab}$ satisfies

$$h_{ab;c} = 2g_{ab}\phi_c + g_{ac}\phi_b + g_{bc}\phi_a \tag{2}$$

for some smooth closed 1-form on $M$ with local components $\phi_a$. Thus $\phi_a$ is locally gradient and will, where appropriate, be written as $\phi_a = \phi_{,a}$ for some function $\phi$ on some open subset of $M$. If $X$ is a projective vector field and $\phi_{a;b} = 0$ then $X$ is called a special projective vector field on $M$. If $h_{ab;c} = 0$ on $M$ is, from (2), equivalent to $\phi_a$ being zero on $M$ and is, in turn equivalent to $X$ being an affine vector field on $M$ (so that the local diffeomorphisms $\psi_t$ preserve not only geodesics but also their affine parameters). If $X$ is projective but not affine then it is called proper projective [2,3]. Further if $X$ is affine and $h_{ab} = 2cg_{ab}, c \in R$ then X is homothetic (otherwise proper affine). If $X$ is homothetic and $c \neq 0$ it is proper homothetic while $c = 0$ it is Killing.

## 2. PROJECTIVE SYMMETRY

If $X$ be a projective vector field on $M$. Then from (1) and (2) [1]

$$L_X R^a{}_{bcd} = \delta^a_d \phi_{b;c} - \delta^a_c \phi_{a;b}, \quad L_X R_{ab} = -3\phi_{a;b}.$$

Also the Ricci identity on $h$ gives

$$h_{ae} R^e{}_{bcd} + h_{be} R^e{}_{acd} = g_{ac}\phi_{b;d} - g_{ad}\phi_{b;c} + g_{bc}\phi_{a;d} - g_{bd}\phi_{a;c}.$$

Let $X$ be a projective vector field on $M$ so that (1) and (2) holds and let $F$ be a real curvature eigenbivector at $p \in M$ with eigenvalue $\lambda \in R$ (so that $R^{ab}{}_{cd} F^{cd} = \lambda F^{ab}$ at $p$) then at $p$ one has [4]

$$P_{ac} F^c{}_b + P_{bc} F^c{}_a = 0 \qquad (P_{ab} = \lambda h_{ab} + 2\phi_{a;b}) \tag{3}$$

Equation (3) gives a relation between $F^a{}_b$ and $P_{ab}$ (a second order symmetric tensor) at $p$ and reflects the close connection between $h_{ab}, \phi_{a;b}$ and the algebraic structure of the curvature at $p$. If $F$ is simple then the blade of $F$ (a two dimensional subspace of $T_pM$) consists of eigenvectors of $P$ with same eigenvalue. Similarly, if $F$ is non-simple then it has two well defined orthogonal timelike and spacelike blades at $p$ each of which consists of eigenvectors of $P$ with same eigenvalue [5].

## 2.1 Existence of Projective vector field in non flat conformally flat cylindrically symmetric static space-times

Consider a cylindrically symmetric static space-time in the usual coordinate system $(t,r,\theta,\phi)$ (labeled by $(x^0, x^1, x^2, x^3)$, respectively) with line element [6]

$$ds^2 = -e^{v(r)}dt^2 + dr^2 + e^{u(r)}d\theta^2 + e^{w(r)}d\phi^2. \qquad (4)$$

Since we are interested in those cases when the above space-time (4) is become conformally flat but non flat, It follows from [7,8] there exists only one possibility which is:

(P1)     $v(r) = u(r) = w(r)$

**Case P1**

In this case the above space-times becomes

$$ds^2 = -e^{v(r)}dt^2 + dr^2 + e^{v(r)}(d\theta^2 + d\phi^2). \qquad (5)$$

The above space-time (6) admits six independent Killing vector fields, which are

$$\frac{\partial}{\partial t},\ \frac{\partial}{\partial \theta},\ \frac{\partial}{\partial \phi},\ \theta\frac{\partial}{\partial \phi} - \phi\frac{\partial}{\partial \theta},\ \theta\frac{\partial}{\partial t} + t\frac{\partial}{\partial \theta},\ \phi\frac{\partial}{\partial t} + t\frac{\partial}{\partial \phi}.$$

These six Killing vector fields are clearly tangent to the family of three dimensional timelike hypersurfaces of constant $r$. Consequently, these hypersurfaces are constant (zero) curvature. The Ricci tensor Segre of the above space-time is $\{(1,11)1\}$ or $\{(1,111)\}$. If the Segre is $\{(1,111)\}$ then the space-time is of constant curvature and the projective vector fields are given in [1]. Here it is assumed that the space-time is not constant curvature. The non-zero independent components of the Riemann tensor are

$$R^{21}{}_{21} = R^{31}{}_{31} = R^{10}{}_{10} = -\frac{1}{4}(2v'' + v'^2) \equiv \beta_2,$$

$$R^{30}{}_{30} = R^{20}{}_{20} = R^{32}{}_{32} = -\frac{1}{4}v'^2 \equiv \beta_1.$$

(6)

Writing the curvature tensor with the components $R^{ab}{}_{cd}$ at $p$ as a $6 \times 6$ matrix in a well known way [9]

$$R^{ab}{}_{cd} = diag(\beta_2, \beta_1, \beta_1, \beta_2, \beta_2, \beta_1)$$

where $\beta_1$ and $\beta_2$ are real functions of $r$ only and where the 6-dimensional labelling is in the order $01, 02, 03, 12, 13, 23$ with $x^0 = t$. Here, at $p \in M$ one may choose a tetrad $(t, r, \theta, \phi)$ satisfying $-t^a t_a = r^a r_a = \theta^a \theta_a = \phi^a \phi_a = 1$ (with all others inner products zero) such that the eigenbivector of the curvature tensor at $p$ are all simple with blades spanned by the vector pairs $(t, r), (r, \theta), (r, \phi)$ each with eigenvalue $\beta_2(p)$ and $(t, \theta), (t, \phi), (\theta, \phi)$ each with eigenvalue $\beta_1(p)$. Here we are considering the open subregion where $\beta_2$ and $\beta_1$ are nowhere equal (if $\beta_1 = \beta_2$ than it follws from (6) the above space-time (5) becomes constant curvature which gives contradiction to our assumtion. So $\beta_1 \neq \beta_2$) and $\beta_2 \neq 0$ (if $\beta_2 = 0$ than the rank of the $6X6$ Riemann matrix become three and it follows from [12] no proper projective will exist. So $\beta_2 \neq 0$). Thus, at $p$ the tensor $P_{ab} = \beta_2 h_{ab} + 2\psi_{a;b}$ has eigenvectors $t, r, \theta, \phi$ with same eigenvalue, say, $\delta_1$ and $P_{ab} = \beta_1 h_{ab} + 2\psi_{a;b}$ has eigenvectors $t, \theta, \phi$ with same eigenvalue, say, $\delta_2$. Hence on $M$ one has after use of the completeness relation

$$\beta_2 h_{ab} + 2\psi_{a;b} = \delta_1 g_{ab}, \quad \beta_1 h_{ab} + 2\psi_{a;b} = \delta_2 g_{ab} + \delta_4 r_a r_b \quad (7)$$

where $\delta_1$, $\delta_2$ and $\delta_4$ are some real functions on $M$. Since $\beta_2 \neq \beta_1$ then it follows from (7) that

$$h_{ab} = C g_{ab} + D r_a r_b, \quad \psi_{a;b} = F g_{ab} + F r_a r_b \quad (8)$$

for some real functions $C, D, E$ and $F$ on $M$. Next one substitutes the first equation of (8) in (2) and contracts the resulting expression first with $t^a \theta^b$ and then $t^a \phi^b$ to get $\psi_a x^a = \psi_a \theta^a = \psi_a \phi^a = 0$ and so $\psi_a = \eta r_a$ for some function $\eta$.

The same expression contracted with $t^a t^b$ gives $C_c = 2\psi_c \Rightarrow C = C(r)$. Now again the same expression contracted with $r^a r^b$ and using the above information gives $D_c = 2\eta\, r_c$ and hence $D = D(r)$. Consider the equation $\psi_a = \eta\, r_a$ and after taking the covariant derivative we get $\psi_{a;b} = \eta\, r_{a;b} + \eta_b\, r_a$. Next consider the second equation of (8) and use $\psi_{a;b} = \eta\, r_{a;b} + \eta_b\, r_a$ and then contract with $r^a$ to get $\eta_a \propto r_a$ so that $\eta = \eta(r)$. Consider the first equation of (8) and use (5) one has the following non-zero components of $h_{ab}$

$$h_{00} = -Ce^\nu,\ h_{11} = (C+D),\ h_{22} = Ce^\nu\ \text{and}\ h_{33} = Ce^\nu. \tag{9}$$

Now we are interested in finding projective vector fields by using the following relation

$$L_X g_{ab} = h_{ab}. \tag{10}$$

Using equation (9) and (5) in (10) and writing out explicitly we get

$$\nu' X^1 + 2 X^0_{,0} = C \tag{11}$$

$$X^1_{,0} - e^\nu X^0_{,1} = 0 \tag{12}$$

$$X^2_{,0} - X^0_{,2} = 0 \tag{13}$$

$$X^3_{,0} - X^0_{,3} = 0 \tag{14}$$

$$X^1_{,1} = \frac{1}{2}(C+D) \tag{15}$$

$$e^\nu X^2_{,1} + X^1_{,2} = 0 \tag{16}$$

$$e^\nu X^3_{,1} + X^1_{,3} = 0 \tag{17}$$

$$\nu' X^1 + 2 X^2_{,2} = C \tag{18}$$

$$X^3_{,2} + X^2_{,3} = 0 \tag{19}$$

$$\nu' X^1 + 2 X^3_{,3} = C. \tag{20}$$

Equations (15), (16), (17) and (12) give

$$X^1 = \frac{1}{2}\int (C+D)dr + A^1(t,\theta,\phi)$$
$$X^2 = -A^1_\theta(t,\theta,\phi)\int e^{-v}dr + A^2(t,\theta,\phi)$$
$$X^3 = -A^1_\phi(t,\theta,\phi)\int e^{-v}dr + A^3(t,\theta,\phi) \quad (21)$$
$$X^0 = A^1_t(t,\theta,\phi)\int e^{-v}dr + A^4(t,\theta,\phi)$$

where $A^1(t,\theta,\phi)$, $A^2(t,\theta,\phi)$, $A^3(t,\theta,\phi)$ and $A^4(t,\theta,\phi)$ are functions of integration. In order to determine $A^1(t,\theta,\phi)$, $A^2(t,\theta,\phi)$, $A^3(t,\theta,\phi)$ and $A^4(t,\theta,\phi)$ we need to integrate the remaining six equations. To avoid details, here we will present only the result. The solution of the equations (11) – (20) is

$$X^0 = ta + \theta c_1 + \phi c_2 + c_3, \quad X^1 = \frac{1}{2}\int (C+D)dr + b,$$
$$X^2 = \theta a + tc_1 - \phi c_4 + c_5, \quad X^3 = \phi a + tc_2 + \theta c_4 + c_6 \quad (22)$$

provided that

$$\int (C+D)dx + b = \frac{1}{v'}(C - 2a) \quad v' \neq 0,$$

where $a,b,c_1,c_2,c_3,c_4,c_5,c_6 \in R$. After subtracting Killing vector fields from (22) one has

$$X^0 = ta, \quad X^1 = \frac{1}{2}\int (C+D)dr + b, \quad X^2 = \theta a, \quad X^3 = \phi a$$

provided that

$$\int (C+D)dx + b = \frac{1}{v'}(C - 2a) \quad v' \neq 0.$$

Suppose $X = (ta, \rho(r), \theta a, \phi a)$, where $\rho(r) = \frac{1}{2}\int (C+D)dr + b$ and $\rho(r) = \frac{1}{v'}(C-2a)$. The vector field $X$ is then projective if it satisfies (2). So, using the above information in (2) gives

$$v'\rho' - v'(a + \frac{1}{2}\rho v') = \frac{1}{2}(v''\rho + v'\rho'), \quad \rho'' = v''\rho + v'\rho' \quad (23)$$

and also $\psi_a = \rho'' r_a$. A particular solution of (23) is

$$\rho = a_1 e^r - 2a, \quad v = r + a_2 \quad (24)$$

where $a_1, a_2 \in R (a_1 \neq 0)$ and $C = D = a_1 e^r$. *Thus the space-time (5) admits a proper projective vector field, for the special choice of v as given in (24).*

## 2.2 Existence of Projective vector field in non flat conformally flat Bianchi type I space-times

Consider a Bianchi type-1 space-time in the usual coordinate system $(t, x, y, z)$ (labeled by $(x^0, x^1, x^2, x^3)$, respectively) with line [10]

$$ds^2 = -dt^2 + k(t)dx^2 + h(t)dy^2 + f(t)dz^2. \tag{25}$$

The above space-time admits three linearly independent killing vector fields, which are

$$\frac{\partial}{\partial x}, \frac{\partial}{\partial y}, \frac{\partial}{\partial z}.$$

Since we are interested in those cases when the above space-time (25) is become conformally flat but non flat, It follows from [7,10] there exists only one possibility which is:

(P2)   $k(t) = h(t) = f(t)$

**Case P2**

In this case the above space-times becomes

$$ds^2 = -dt^2 + k(t)(dx^2 + dy^2 + dz^2) \tag{26}$$

and it admits six independent Killing vector fields, which are

$$y\frac{\partial}{\partial x} - x\frac{\partial}{\partial y}, \; z\frac{\partial}{\partial x} - x\frac{\partial}{\partial z}, \; y\frac{\partial}{\partial z} - z\frac{\partial}{\partial y}, \; \frac{\partial}{\partial x}, \frac{\partial}{\partial y}, \frac{\partial}{\partial z}.$$

These six Killing vector fields are clearly tangent to the family of three dimensional timelike hypersurfaces of constant *t*. Consequently, these hypersurfaces are constant (zero) curvature. The Segre types of the above space-time are {1, (111)} or {(1,111)}. If the Segre is {(1,111)} then the space-time is of constant curvature and the projective vector fields are given in [1]. Here it is assumed that the space-time is not of constant curvature. The proper projective vector fields for the above space-time (26) are also available in [2, 11, 13]. The non-zero independent components of the Riemann curvature tensors are

$$R^{01}{}_{01} = R^{02}{}_{02} = R^{03}{}_{03} = \frac{1}{k}\left(\frac{\ddot{k}}{2} - \frac{\dot{k}^2}{4k}\right) \equiv A$$

$$R^{12}{}_{12} = R^{13}{}_{13} = R^{32}{}_{32} = \frac{1}{k}\left(\frac{\dot{k}^2}{4k}\right) \equiv B. \tag{27}$$

Writing the curvature tensor with the components $R^{ab}{}_{cd}$ at $p$ as a $6 \times 6$ matrix in a well known way [9]

$$R^{ab}{}_{cd} = diag(A, A, A, B, B, B)$$

where $A$ and $B$ are real functions of $t$ only and where the 6-dimensional labelling is in the order $01, 02, 03, 12, 13, 23$ with $x^0 = t$. Here, at $p \in M$ one may choose a tetrad $(t, x, y, z)$ satisfying $-t^a t_a = x^a x_a = y^a y_a = z^a z_a = 1$ (with all otherr inner products zero) such that the eigenbivectors of the curvature tensor at $p$ are all simple with blades spanned by the vector pairs $(t, x), (t, y), (t, z)$ each with eigen value $A(p)$ and $(x, y), (x, z), (y, z)$ each with eigenvalue $B(p)$. Here we are considering the open subregion where $A$ and $B$ are nowhere equal (if $A = B$ than it follws from (27) the above space-time (26) becomes constant curvature which gives contradiction to our assumtion. Hence $A \neq B$) and $A \neq 0$ (if $A = 0$ than the rank of the $6X6$ Riemann matrix become three and it follows from [12] no proper projective will exist. Hence $A \neq 0$). Thus, at $p$ the tensor $P_{ab} = Ah_{ab} + 2\psi_{a;b}$ has eigenvectors $t$, $x$, $y$ and $z$ with same eigenvalue, say, $\gamma_1$ and $P_{ab} = Bh_{ab} + 2\psi_{a;b}$ has eigenvectors $x$, $y$ and $z$ with same eigenvalue, say, $\gamma_2$. First consider the equation $P_{ab} = Ah_{ab} + 2\psi_{a;b}$, where $P_{ab}$ is a second order symmetric tensor with eigenvectors $t$, $x$, $y$ and z with same eigenvvalue $\gamma_1$. The Segre type of $P_{ab}$ is $\{(1,111)\}$, and $P_{ab} = \gamma_1 g_{ab}$. Substituting back, we get $Ah_{ab} + 2\psi_{a;b} = \gamma_1 g_{ab}$. Now consider $P_{ab} = Bh_{ab} + 2\psi_{a;b}$, where $P_{ab}$ is a second order symmetric tensor with eigenvectors $x$, $y$ and $z$ with same eigenvalue, say, $\gamma_2$. The Segre type of $P_{ab}$ is $\{1,(111)\}$ and $P_{ab} = \gamma_2 g_{ab} + \gamma_3 t_a t_b$. Substituting back, we get $Bh_{ab} + 2\psi_{a;b} = \gamma_2 g_{ab} + \gamma_3 t_a t_b$. Hence on $M$ one has

$$Ah_{ab} + 2\psi_{a;b} = \gamma_1 g_{ab}, \qquad Bh_{ab} + 2\psi_{a;b} = \gamma_2 g_{ab} + \gamma_3 t_a t_b \qquad (28)$$

where $\gamma_1, \gamma_2$ and $\gamma_3$ are some real functions on $M$. Since $A \neq B$ then it follows from equation (28) that

$$h_{ab} = \beta g_{ab} + \alpha t_a t_b, \qquad \psi_{a;b} = E g_{ab} + F t_a t_b \qquad (29)$$

for some real functions $\alpha, \beta, E$ and $F$ on $M$. Now one substitutes the first equation of (29) in (2) and contracts the resulting expression with $x^a y^b$ and then with $x^a z^b$ to get $\psi_a x^a = \psi_a y^a = \psi_a z^a = 0$ and one has $\psi_a = \xi t_a$ for some function $\xi$. The same expression contracted with $t^a t^b$ then show that $(\alpha - \beta)_{,a} = -4\xi t_a$ and so $(\alpha - \beta)$ is a function of $t$ only. Now again contract the same expression with $x^a x^b$ one finds $\beta_c = 2\xi t_c$ which implies $\beta_a x^a = \beta_a y^a = \beta_a z^a = 0 \Rightarrow \beta = \beta(t)$. Substituting back to get $\alpha_c = -2\xi t_c$ and hence $\alpha = \alpha(t)$. Now consider the second equation of (29) and use $\psi_{a;b} = \xi_b t_a + \xi t_{a;b}$ and contract with $t^a$. One can easily find that $\xi = \xi(t)$. Consider the first equation of (29) and use (26) one has the following non zero components of $h_{ab}$

$$h_{00} = (\alpha - \beta), \quad h_{11} = \beta k, \quad h_{22} = \beta k \text{ and } h_{33} = \beta k, \qquad (30)$$

where $\alpha = \alpha(t)$, $\beta = \beta(t)$ and $\alpha - \beta = (\alpha - \beta)(t)$. Now we are interested in finding projective vector fields by using the relation (10). Writing out equation (10) explicitly and using (26) and (30), we get

$$X^0{}_{,0} = \frac{1}{2}(\beta - \alpha) \qquad (31)$$

$$kX^1{}_{,0} - X^0{}_{,1} = 0 \qquad (32)$$

$$kX^2{}_{,0} - X^0{}_{,2} = 0 \qquad (33)$$

$$kX^3{}_{,0} - X^0{}_{,3} = 0 \qquad (34)$$

$$\frac{1}{2}\dot{k}X^0 + kX^1{}_{,1} = \frac{1}{2}\beta k \tag{35}$$

$$X^2{}_{,1} + X^1{}_{,2} = 0 \tag{36}$$

$$X^3{}_{,1} + X^1{}_{,3} = 0 \tag{37}$$

$$\frac{1}{2}\dot{k}X^0 + kX^2{}_{,2} = \frac{1}{2}\beta k \tag{38}$$

$$X^3{}_{,2} + X^2{}_{,3} = 0 \tag{39}$$

$$\frac{1}{2}\dot{k}X^0 + kX^3{}_{,3} = \frac{1}{2}\beta k. \tag{40}$$

Equations (31), (32), (33) and (34), give

$$\left.\begin{aligned}
X^0 &= \frac{1}{2}\int(\beta-\alpha)dt + A^1(x,y,z) \\
X^1 &= A^1_x(x,y,z)\int\frac{1}{k}dt + A^2(x,y,z) \\
X^2 &= A^1_y(x,y,z)\int\frac{1}{k}dt + A^3(x,y,z) \\
X^3 &= A^1_z(x,y,z)\int\frac{1}{k}dt + A^4(x,y,z)
\end{aligned}\right\}, \tag{41}$$

where $A^1(x,y,z), A^2(x,y,z), A^3(x,y,z)$ and $A^4(x,y,z)$ are functions of integration. In order to determine $A^1(x,y,z), A^2(x,y,z), A^3(x,y,z)$ and $A^4(x,y,z)$ we need to integrate the remaining six equations. To avoid lengthy calculations, here we will present only the result. The solution of the equations (31) – (40) is

$$\left.\begin{aligned}
X^0 &= \frac{1}{2}\int(\beta-\alpha)dt + c_8 \\
X^1 &= xc_1 - yc^5 + zc^6 + c^7 \\
X^2 &= yc_1 + xc^5 - zc^8 + c^{10} \\
X^3 &= zc_1 - xc^6 + yc^8 + c^9
\end{aligned}\right\} \tag{42}$$

provided that

$$\frac{1}{2}\int(\beta-\alpha)dt + c_8 = \frac{k}{\dot{k}}(\beta - 2c_1) \qquad \dot{k} \neq 0,$$

where $c_1, c_8, c^5, c^6, c^7, c^8, c^9, c^{10} \in R$. After subtracting Killing vector fields from (42) one has

$$X^0 = \frac{1}{2}\int(\beta - \alpha)dt + c_8, \quad X^1 = xc_1, \quad X^2 = yc_1, \quad X^3 = zc_1$$

provided that

$$\frac{1}{2}\int(\beta - \alpha)dt + c_8 = \frac{k}{\dot{k}}(\beta - 2c_1) \qquad \dot{k} \neq 0,$$

Suppose $X = (\eta(t), xc_1, yc_1, zc_1)$, where $\eta(t) = \frac{1}{2}\int(\beta - \alpha)dt + c_8$ and

$\frac{k}{\dot{k}}(\beta - 2c^1) = \eta(t)$. The vector field $X$ is said to be projective if it satisfies (2).

So using the above information in (2) gives

$$\frac{\ddot{\eta}}{2} = \frac{\dot{k}}{k}\dot{\eta} - \frac{\dot{k}}{k^2}\left(k c^1 + \frac{1}{2}\dot{k}\eta\right)$$

$$\frac{\ddot{\eta}}{2} = \frac{1}{2k}(\ddot{k}\eta + \dot{k}\dot{\eta}) - \frac{\dot{k}^2}{2k^2}\eta$$

(43)

and $\psi_a = \ddot{\eta} t_a$. A particular solutions of (43) are

$$\eta = k = \frac{I}{F}e^{Ft+FG} - \frac{2c_1 I}{F}$$

(44)

$$k = Le^t, \quad \eta = Ne^t - D,$$

(45)

where $F, G, I, L, N, R \in R (F \neq 0)$. *Thus the space-time (25) admits a proper projective vector field, for the special choice of $k$ as given in (44) and (45).*